\documentclass[a4paper]{article}

\usepackage{INTERSPEECH2020}
\usepackage{epstopdf}

\title{ASR-Free Pronunciation Assessment}
\name{Sitong Cheng, Zhixin Liu, Lantian Li, Zhiyuan Tang, Dong Wang, Thomas Fang Zheng}
\address{
  Center for Speech and Language Technologies, Tsinghua University, China}
\email{wangdong99@mails.tsinghua.edu.cn}

\begin{document}

\maketitle
\begin{abstract}
Most of the pronunciation assessment methods are based on local features derived from automatic speech recognition (ASR), e.g.,
the Goodness of Pronunciation (GOP) score. In this paper, we investigate an ASR-free scoring approach that is derived from
the marginal distribution of raw speech signals.
The hypothesis is that even if we have no knowledge of the language (so cannot recognize the phones/words),
we can still tell how good a pronunciation is, by comparatively listening to some speech data from the target language.
Our analysis shows that this new scoring approach provides an interesting correction for the
phone-competition problem of GOP.
Experimental results on the ERJ dataset demonstrated that combining the ASR-free score and GOP can achieve better performance than the GOP baseline.

\end{abstract}

\section{Introduction}

Automatic pronunciation assessment plays a key role in Computer Assisted Language Learning (CALL). The research in this
field dates back to $90$'s in the last century~\cite{neumeyer1996automatic,franco1997automatic}, and has gained much popularity in
second language education, e.g., ~\cite{zechner2009automatic,eskenazi2009overview,Cheng2018,Miwardelli2019,Yarra2019}.

Most of existing assessment approaches are based on automatic speech recognition (ASR). A simple approach is to count the
words correctly recognized per minute (WCPM). A more careful assessment is often conducted in three steps:
(1) Employ an ASR system to segment speech signals into pronunciation units, e.g., phones or words; (2) Compute local or
global features based on the segmentation; (3) Derive proficiency scores from the local/global features. The features that have been
used can be categorized into two classes: phonetic features and prosodic features. Phonetic features reflect the
quality of the pronunciation of single phones (or words). The most successful phonetic features are the phone-level likelihood or
posterior~\cite{neumeyer1996automatic,franco1997automatic}, others include phone duration, formants, articulation class,
phonetic distance, etc~\cite{Kyriakopoulos2019,Jenne2019}.
Prosodic features reflect the intonation, stress or fluency. The most popular prosodic features include the pitch and intensity
contour, rate of speech (phones per minute), and duration of silence~\cite{Truong2018,Graham2019}.
%Clearly, the prosodic features are generally suprasegmental.
Deriving pronunciation scores from these features can be accomplished by a simple average (e.g., if the features are phone-level posteriors),
or a complex regression model (e.g., if the features are pitch and intensity contour) such as Gaussian process~\cite{knill2015automatically}.
More technical details for pronunciation assessment can be found in Witt's review paper~\cite{witt2012automatic}.

The Goodness of Pronunciation (GOP) is perhaps the most popular features for pronunciation assessment~\cite{witt1999use}.
GOP is based on the posterior probability on the correct phone, given the speech segment of that phone. This is formulated by:

\begin{equation}
\label{eq:gop}
GOP = \frac{1}{M} \sum_i^{M} \ln  p(q_i|o_i),
\end{equation}

\noindent where $q_i$ is the $i$-th phone in the speech segment, and $o_i$ is the corresponding speech segment,
and $M$ is the total number of phones in the speech segment.

Early research computes GOP using Gaussian mixture model-Hidden Markov model (GMM-HMM)~\cite{witt1999use}, and recent study
usually uses acoustic models based on deep neural networks (DNNs)~\cite{Sudhakara2019,hu2013new,hu2015improved,huang2017transfer}.
The DNN-based acoustic modeling offers much more robustness against ambient complexity and speaker variation, which in turn leads to better phone
segmentation and posterior estimation. Due to the high performance and simple computation, we use the DNN-based GOP as the baseline.

In spite of the prominent success of the existing approaches, in particular those based on GOP, the present studies heavily rely on
ASR (for segmentation and posterior estimation). For L2 learners, especially those in the primary stage,
the pronunciations tend to be significantly different from native pronunciations, resulting in high alignment/recognition errors and low quality of
pronunciation assessment~\cite{Knill2018}. This performance reduction could be more severe for languages without strong ASR systems.

This inspired us to investigate an ASR-free scoring approach for pronunciation assessment, by which the assessment is not based on ASR and
so would not be impacted by the ASR performance.
Interestingly, human beings seem to use this way to judge the proficiency of a pronunciation.
Our experience is that even we have no knowledge of the language (so cannot recognize the phones/words),
we can still tell how good a pronunciation is, by listening to some native speech samples from the target language.
In this paper, we present an ASR-free scoring approach that does not rely on ASR but is based on a generative model $p(\mathbf{o})$.
Our theoretical study shows that this score offers an interesting correction for the phone-competition problem of GOP,
and empirical study demonstrates that better performance can be achieved when combining the ASR-free score and GOP.

The reset of the paper is organized as follows. Section~\ref{sec:gop} analyzes the potential problem of GOP,
and Section~\ref{sec:method} presents the ASR-free scoring approach.
Section~\ref{sec:exp} presents the experimental results on the ERJ dataset,
and Section~\ref{sec:con} concludes the paper.

\section{GOP is not perfect}
\label{sec:gop}

Given a phone sequence $\mathbf{q}$ and the corresponding speech signal $\mathbf{o}$, and assume the alignment is perfect, then the averaged conditional probability
will be the theoretically sound measurement for testing how well a test speech matches the training speech:

\begin{equation}
p(\mathbf{o}|\mathbf{q}) = \frac{1}{M} \sum_i \ln p(o_i|q_i).
\label{eq:conditional}
\end{equation}

\noindent Considering the ideal case where the model is trained on native speech, and the only difference between the test speech and
the training speech is pronunciation proficiency, then the conditional above will be perfect for pronunciation assessment.
In real situations, however, $p(\mathbf{o}|\mathbf{q})$ may be varied by numerous factors that are
not related to pronunciation proficiency, e.g., noise, channel, speaking rate/volume and speaker trait.
This sensitivity to multiple variations means that the conditional is not reliable for pronunciation assessment
in real situations.

We can rewrite the conditional into another form:

\[
p(\mathbf{o}|\mathbf{q}) = \frac{1}{M}  \sum_{i=1}^M \ln \frac{p(q_i|o_i)p(o_i)}{p(q_i)}.
\]
\noindent If we ignore the prior $p(q_i)$, the conditional involves two parts: the posterior $p(q_i|o_i)$ and the marginal $p(o_i)$.
The marginal part inherits the property of the conditional and is sensitive to all variations including pronunciation proficiency.
The posterior part, which is essentially GOP, however, is more robust. It is purely phone discriminative,
and so less sensitive to phone-unrelated variations. This insensitivity
offers an important merit for pronunciation assessment and explains the success of GOP.

%By assuming that native speech is more easily recognized by the native model, one may conclude that the higher posterior
%signifies more native-like pronunciation.
%A potential problem of the GOP score is that the sensitivity to pronunciation proficiency
%could be reduced when it gains the robustness against other variations.

%By assuming that native speech is more easily recognized by the native model, one may conclude that the higher posterior
%signifies more native-like pronunciation.
%In extreme cases, non-native pronunciations may lead to higher posterior rather than lower.

\begin{figure}[t]
  \centering
  \includegraphics[width=0.9\linewidth]{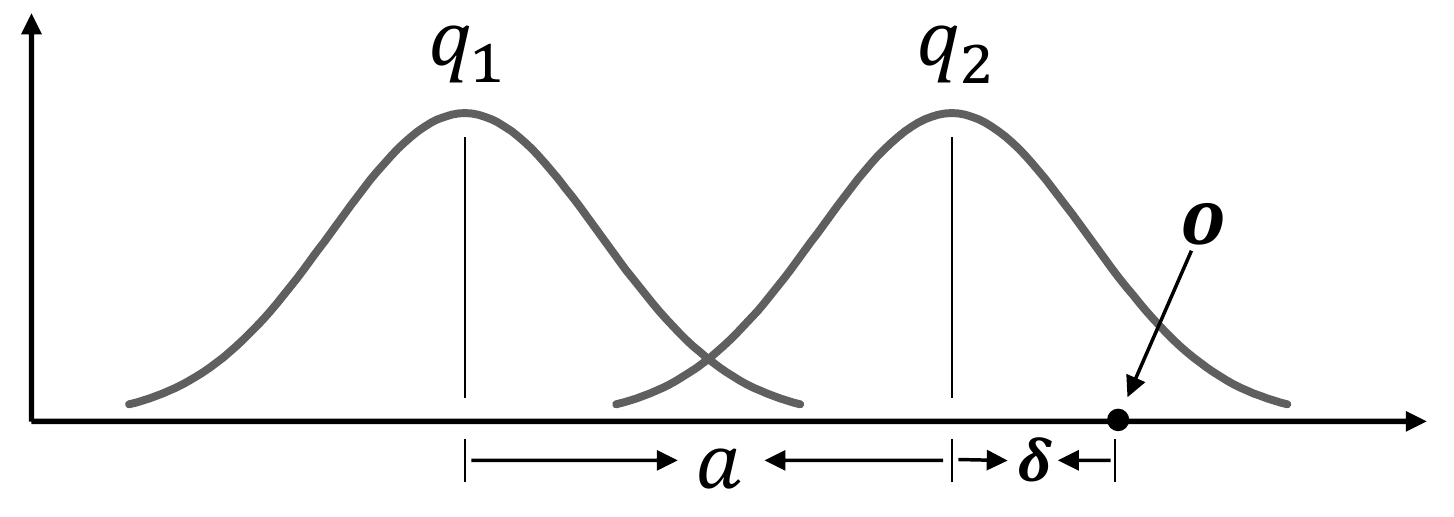}
  \caption{Two conditional distributions for phone $q_1$ and $q_2$. Both are Gaussians with variance $0.5$ and the distance of their means is $a$.
  A small change $\delta$ on the pronunciation in $q_2$ may lead to either increase or decrease on the posterior, depending on the sign of $\delta$. }
  \label{fig:two-gaussian}
\end{figure}

A potential problem of the GOP score is that there is no guarantee that a worse pronunciation will achieve
a smaller posterior.
Let's design a simulation experiment to discuss this issue.
As shown in Figure~\ref{fig:two-gaussian}, we assume two phones $q_1$ and $q_2$ are two one-dimensional Gaussians whose variances are both $0.5$,
and the distance of their means is $a$. At the mean of $q_2$, which can be regarded as a perfect pronunciation of $q_2$,
the posterior on $q_2$ is $p(q_2|o)=\frac{1}{1+e^{-a^2}}$.
Assume a non-native speaker pronounce $q_2$ at a position $\mathbf{o}$, and the shift from the mean of $q_2$ to $\mathbf{o}$ is $\delta$.
The posterior on $q_2$ given $\mathbf{o}$ can be computed as follows:

\[
p(q_2|\mathbf{o}) = \frac{e^{-\delta^2}}{e^{-\delta^2} + e^{-(a + \delta)^2}  } = \frac{1}{1 + e^{-(a^2+2a\delta)}}.
\]

\noindent It can be seen that the change of the posterior depends on the sign of $\delta$:
if $\delta > 0$, the posterior essentially increases. This means that a non-native speaker obtains a better
GOP than a native speaker. It clearly demonstrates that GOP is not a perfect score, at least in theory.

\section{Proposed methods}
\label{sec:method}

\subsection{ASR-free scoring}

%Since the conditional probability $p(o|q)$ in Eq.~\ref{eq:conditional} does not has the problem of the GOP score has, it is easy to conjecture that this problem can be
%amended by taking the marginal part $p(o)$ back into account.

The undesired behavior of GOP is essentially caused by the \emph{competition between phones}.
Recalling that the conditional $p(\mathbf{o}|\mathbf{q})$ in Eq.~(\ref{eq:conditional}) is a perfect assessment
and does not suffer from this problem, we conjecture that it is the marginal part $p(\mathbf{o})$ that solves the
phone competition. Therefore, we argue that $p(\mathbf{o})$ should be involved in the assessment, rather than simply discarded.
Since $p(\mathbf{o})$ concerns neither phones nor words, it is called an \textbf{ASR-free score}.

However, simply multiplying $p(\mathbf{q}|\mathbf{o})$ by $p(\mathbf{o})$ as in Eq.~(\ref{eq:conditional}) does not work,
as $p(\mathbf{o})$ is quite noisy. To demonstrate this argument,
we trained a GMM model using the WSJ native English dataset and computed the Pearson correlation coefficient (PCC)
between the marginal distribution $p(\mathbf{o})$ and the human-labelled scores on the ERJ (English read by Japanese) dataset,
and found that the PCC is nearly zero or even negative. More details will be presented in Section~\ref{sec:exp}.

In order to employ $p(\mathbf{o})$ but reduce the noise,
a discriminative model can be used to discover the factors $\mathbf{z}_s$ in $\mathbf{o}$ that are mostly related to pronunciation proficiency,
and then build $p(\mathbf{z}_s)$ rather than $p(\mathbf{o})$.
This approach, however, requires the native speech being labelled by pronunciation proficiency, which is almost impossible.
There is a more direct and parsimonious way: since our goal is to assess the pronunciation, we can
build a \textbf{conditional model} $p(s|\mathbf{o})$ that not only selects the factor $\mathbf{z}_s$, but also produces the assessment score $s$ at the same time.

\begin{figure}[t]
  \centering
  \includegraphics[width=0.7\linewidth]{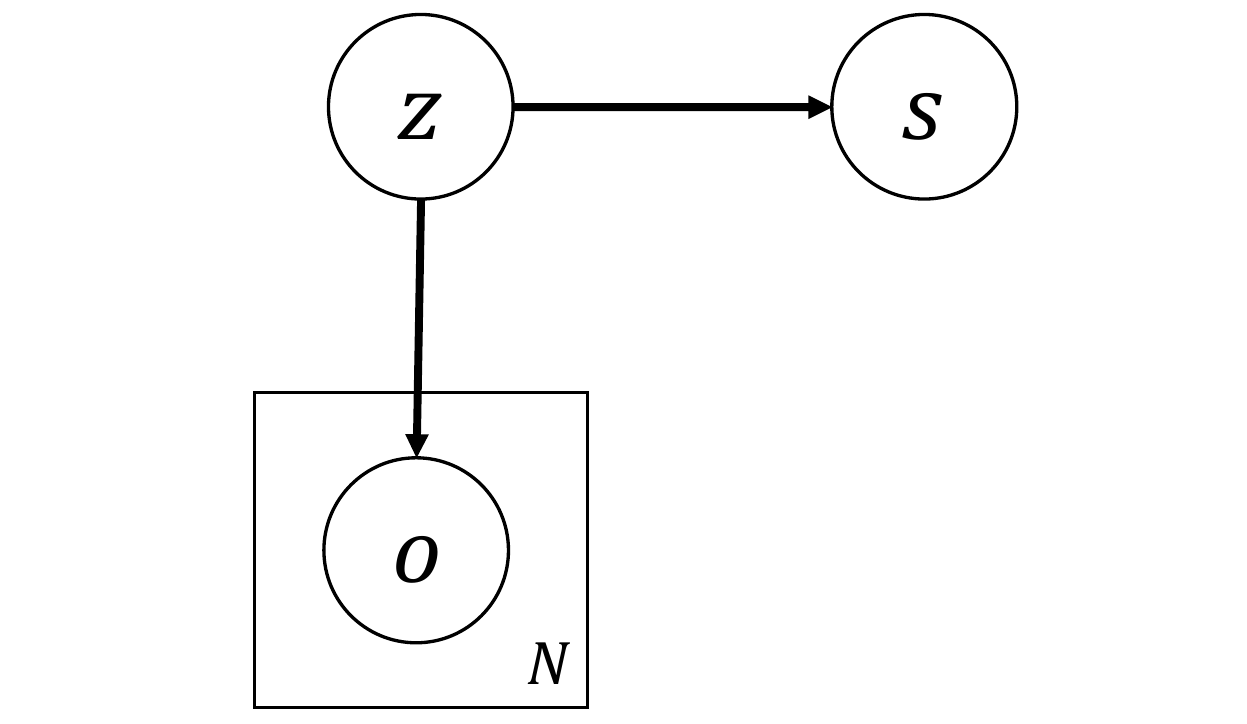}
  \caption{The graphical model for ASR-free score estimation, where $\mathbf{o}$ is the speech segment that consists of $N$ frames, $\mathbf{z}$ is the
  utterance-level representation of $\mathbf{o}$, $s$ is the assessment score.}
  \label{fig:graph}
\end{figure}

We therefore propose a probabilistic model shown in Figure~\ref{fig:graph}.
Firstly, we build a \textbf{marginal model} $p(\mathbf{o})$ that describes all the variations in $\mathbf{o}$. This model, however, is not used to compute the marginal probability;
instead, it is used to infer the utterance-level representation $\mathbf{z}$ for the speech $\mathbf{o}$.
%This step is an unsupervised training and can exploit a large amount of speech data, either native or non-native.
Secondly, we build a \textbf{prediction model} $p(s|\mathbf{z})$ that selects $\mathbf{z}_s$ from $\mathbf{z}$ and produces the assessment score $s$.
This step is a supervised training and a small amount of training
data is sufficient. Combining the two models, we can build the conditional model $p(s|\mathbf{o})$.

\subsection{Marginal model}

There are several ways to build the marginal model $p(\mathbf{o})$, but not all of them can infer the latent representation $\mathbf{z}$ in a compact way.
We consider three models: i-vector, normalization flow (NF) and discriminative normalization flow (DNF).

\subsubsection{i-vector model}

The i-vector model is a mixture of linear Gaussians~\cite{dehak2011front}:

\[
 p(\mathbf{o}) = \sum_k \pi_k N(\mathbf{o}|\mathbf{T}_k \mathbf{z}, \pmb{\Sigma_k}),
\]
\[
 p(\mathbf{z}) = N(\mathbf{z}; \mathbf{0}, \mathbf{I}),
\]

\noindent  where $k$ indexes the Gaussian component, and the loading matrices $\{\mathbf{T}_k\}$ are low-rank.
This is a full-generative model, and the posterior on $\mathbf{z}$
can be easily computed given a speech signal $\mathbf{o}$. The mean of the posterior $p(\mathbf{z}|\mathbf{o})$ is the i-vector.

The i-vector model possesses several advantages: (1) The model is trained in an unsupervised way and therefore can exploit rich unlabelled data;
(2) The Gaussian assumption makes the training and inference simple; (3) The form of Gaussian mixtures absorbs the impact of
short-time variations (e.g., speech content), which makes the i-vectors represent the long-term variations.
For these reasons, the i-vector
model has been widely employed in various speech processing tasks, such as speaker recognition~\cite{dehak2011front} and
language recognition~\cite{dehak2011language,martinez2011language}.

\subsubsection{Normalization flow}

The i-vector model is a shallow model, which may prevent it from representing complex distributions.
Normalization flow (NF) is a deep generative model~\cite{papamakarios2019normalizing} and can describe more complex distributions.
The foundation of NF is the principle of distribution transformation for continuous variables~\cite{rudin2006real}.
Let a latent variable $\mathbf{z}$ and an observation variable $\mathbf{o}$ be linked by an invertible transform $\mathbf{o} = f(\mathbf{z})$, their probability density has the following relationship:

\[
\ln p(\mathbf{o}) = \ln p(\mathbf{z}) + \ln \Big | \det \frac{ \partial f^{-1}(\mathbf{o})}{\partial \mathbf{o}} \Big |,
\]

\noindent where $f^{-1}(\mathbf{o})$ is the inverse function of $f(\mathbf{z})$. It has been shown that if $f$ is flexible enough,
a simple distribution (a standard Gaussian) can be transformed to a complex distribution.
%Fig.~\ref{fig:change} illustrates how a complex distribution is normalized to a simple distribution by an NF model.
Usually, $f$ is implemented as a composition of a sequence of relatively simple invertible transforms~\cite{tabak2013family}.

%\begin{figure}[htb]
%    \centering
%    \includegraphics[width=1\linewidth]{flow-map-mono.png}
%    \caption{A complex mixture of three Gaussians is transformed to a single Gaussian by an NF model.  The NF used here is a masked autoregressive
%    flow (MAF)~\cite{papamakarios2017masked}, and the distribution on $\mathbf{z}$ is a standard Gaussian.}
%    \label{fig:change}
%\end{figure}

Once the model has been trained, for a speech segment $\mathbf{o}$,
the latent variable $\mathbf{z}$ can be inferred by averaging the image of $o_i$ in the latent space:

\begin{equation}
  \mathbf{z} = \frac{1}{N} \sum_{i=1}^{N} f^{-1}(o_i) = \frac{1}{N} \sum_{i=1}^{N} z_i.
\label{eq:nf}
\end{equation}

\subsubsection{Discriminative NF}

\begin{figure}[htb]
    \centering
    \includegraphics[width=0.9\linewidth]{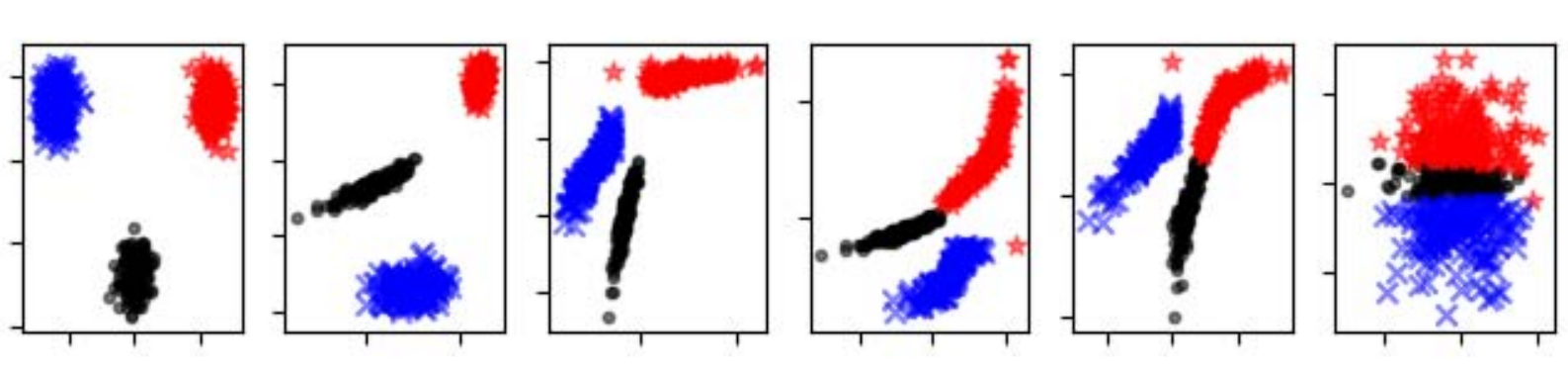}
    \includegraphics[width=0.9\linewidth]{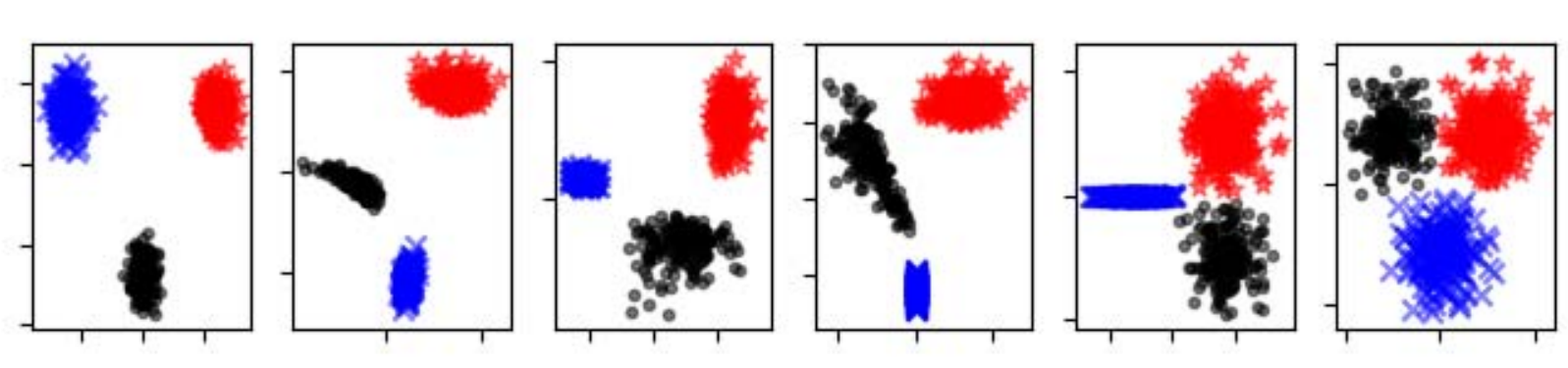}
    \caption{Vanilla NF (top) pulls all classes together in the latent space, while DNF (bottom) keeps data from different classes separated.}
    \label{fig:congress}
\end{figure}

The vanilla NF model optimizes the distribution of the training data without considering the class labels.
%, i.e. the marginal distribution.
This means that data from different classes tend to congest together in the latent space, as shown in the top row of Figure~\ref{fig:congress}.
This is not a good property for downstream applications that require discrimination within the latent space, e.g., pronunciation proficiency.

Recently, the authors proposed a discriminative NF (DNF) model to deal with this problem~\cite{cai2020deep}.
The main advantage of DNF is that it allows each class to have its own Gaussian prior, i.e.
all the priors share the same covariance but possess different means, formulated as follows:

\[
p_s(\mathbf{z}) =  N(\mathbf{z}; \pmb{\mu}_s, \pmb{\Sigma}),
\]
\noindent where $s$ is the class label. By setting class-specific means, different classes will be separated from each other in the latent space, as shown in the bottom row of Figure~\ref{fig:congress}.
If we treat data with different pronunciation proficiency as different classes and train DNF as mentioned above, it will be possible to establish a latent space which is more discriminative for
pronunciation assessment.
Once the model has been trained, the same averaging approach Eq.~(\ref{eq:nf}) as in NF can be used to derive the utterance-level representation
$\mathbf{z}$.

\subsection{Prediction model}

Modeling the prediction probability $p(s|\mathbf{z})$ is easy, if $\mathbf{z}$ has been derived from the marginal model $p(\mathbf{o})$.
Linear regression is the most popular choice, but we found support vector regression (SVR) often produces better results.
Compared to linear regression, SVR chooses a few examples that hold the best predictive
power as support vectors. It is argued that this model is more robust against data sparsity and data imbalance. With SVR, we predict the
score $s$ directly, which can be regarded as a special form of the prediction distribution $p(s|\mathbf{z})$
where all the probability mass concentrates in a single value.

\subsection{Information fusion}

According to Eq.~(\ref{eq:conditional}), a perfect assessment score should combine the posterior part $p(\mathbf{q}|\mathbf{o})$ (which is the GOP score)
and the marginal part $p(\mathbf{o})$ (which is the ASR-free score).
As we mentioned, $p(\mathbf{o})$ is very noisy and so we model $p(s|\mathbf{o})$ instead.
Since $p(\mathbf{o})$ and $p(s|\mathbf{o})$ model different uncertainties,
simply substitute $p(\mathbf{o})$ for $p(s|\mathbf{o})$ in Eq.~(\ref{eq:conditional}) is not theoretically correct.
There are two approaches to combine $p(\mathbf{q}|\mathbf{o})$ and $p(s|\mathbf{o})$: score fusion and feature fusion.

\subsubsection{Score fusion}

In score fusion, we treat the GOP score $p(\mathbf{q}|\mathbf{o})$ and the ASR-free score $p(s|\mathbf{o})$ as two independent scores,
and interpolate them in the assessment:

\[
s^* =  \lambda p(\mathbf{q}|\mathbf{o}) + (1-\lambda) \arg\max_{s} \{p(s|\mathbf{o})\}.
\]

\noindent Since SVR outputs the optimal $s$, this fusion score reduces to a simple form:

\[
s^* = \lambda p(\mathbf{q}|\mathbf{o}) + (1-\lambda) \gamma(\mathbf{o}),
\]
\noindent where $\gamma(\cdot)$ is the prediction function implemented by SVR.

%Another possible form is:
%
%\[
%s^* = max_{s } \{ p(q|o) p(s|o)^{\lambda}\}
%\]
%
%\noindent Since SVR outputs the best $s$, this fusion score reduces to a simple form:
%
%\[
%s^* =   p(q|o) g_{SVR}(o)^{\lambda}
%\]
%\noindent where $g_{SVR}$ is the prediction function implemented by the SVR model.

\subsubsection{Feature fusion}

In the feature fusion, we treat the GOP score $p(\mathbf{q}|\mathbf{o})$ as a feature and combine it with the
latent representation $\mathbf{z}$, and then build the SVR model. This feature fusion may discover
valuable knowledge in $\mathbf{z}$ that has not been represented by $p(\mathbf{q}|\mathbf{o})$.
It produces good performance in our experiments.

\section{Experiments}
\label{sec:exp}

\subsection{Data}

Two datasets were used in our experiments: WSJ (Wall Street Journal) and ERJ (English Read by Japanese)~\cite{minematsu2004development}.

\noindent \textbf{WSJ}: A native English speech dataset. It contains $37,318$ utterances from $282$ speakers.
The dataset was used to train the DNN-HMM ASR system to build the GOP baseline.

\noindent \textbf{ERJ}: A standard Japanese-speaking-English dataset.
It consists of $2,000$ utterances from $190$ speakers.
Each utterance has $5$ human-labelled pronunciation scores.
There are $5$-scale ratings as the pronunciation proficiency scores.
In our experiments, the dataset was separated into two subsets: \emph{ERJ.Train} and \emph{ERJ.Eval}.
\emph{ERJ.Train} consists of $1,520$ utterances and was used to train the marginal model (i-vector, NF, DNF) and the prediction model (SVR).
\emph{ERJ.Eval} consists of $380$ utterances and was used for performance tests.
The Pearson correlation coefficient (PCC) on \emph{ERJ.Eval} of human-labelled scores is $0.550$.

%\subsection{Evaluation Metric}
%
%Pearson correlation coefficient (PCC) is used to measure the correlation between the predicted scores $X$ and the human-labelled scores $Y$:
%
%\[
%\rho(X,Y) = \frac{cov(X,Y)}{\sigma(X)\sigma(Y)} = \frac{E[(X-\mu_x)(Y-\mu_y)]}{\sigma(X)\sigma(Y)},
%\]
%
%\noindent where $cov(X,Y)$ is the covariance of $X$ and $Y$, $\sigma(X)$ and $\sigma(Y)$ are the standard deviations on $X$ and $Y$, respectively.
%The value of $\rho(X,Y)$ ranges from -$1$ to $1$. The larger $\rho(X,Y)$, the better correlation between $X$ and $Y$.

\subsection{Model Settings}

\noindent \textbf{DNN-HMM}: It was built using the Kaldi toolkit~\cite{povey2011kaldi}, following the WSJ s5 nnet3 recipe.
The DNN structure consists of $6$ time-delay layers, each followed by a P-norm layer that reduces the dimensionality of the activations
from $2,048$ to $256$. The input features are $23$-dimensional Fbanks with the context window of $21$ frames,
and the output layer contains $3,360$ units, corresponding to the number of GMM-HMM senones.
This model was used to create the GOP baseline as Eq.~(\ref{eq:gop}).

\noindent \textbf{GMM}: It was created using the Kaldi toolkit, following the SITW recipe.
The raw features involve $23$-dimensional MFCCs plus the log energy, augmented by first- and second-order derivatives,
resulting in a $72$-dimensional feature vector. The number of Gaussian components is set to $64$.
Once trained, the log-likelihood of each frame $p(o_i)$ can be estimated,
and the utterance-level $p(\mathbf{o})$ can be computed by a simple average.

\noindent \textbf{i-vector}: The data preparation is the same as GMM.
The number of Gaussian components of the UBM is $64$, and the dimensionality of the i-vector is $440$.

\noindent \textbf{NF}: It was trained using the PyTorch toolkit. The NF used here is a RealNVP architecture~\cite{dinh2016density}.
This model has $10$ non-volume preserving (NVP) layers.
The input features are $40$-dimensional Fbanks with the effective context window of $11$ frames,
leading to a $440$-dimensional feature vector.
The Adam optimizer~\cite{kingma2014adam} is used to train the model with the learning rate set to $0.001$.
Similar to GMM, the utterance-level $p(\mathbf{o})$ can be computed.
In addition, the utterance-level representation $\mathbf{z}$ of the speech $\mathbf{o}$ can be derived by Eq.~(\ref{eq:nf}).

\noindent \textbf{DNF}: The data preparation and model structure are the same as NF.
The number of classes is set to $5$ corresponding to the $5$-scale pronunciation proficiency scores in the ERJ dataset.
The utterance-level representation $\mathbf{z}$ can also be derived by Eq.~(\ref{eq:nf}).

\noindent \textbf{SVR}: It was implemented using the Scikit-Learn toolkit with the default SVR configuration.

\subsection{Basic results}

\noindent The basic results on the \emph{ERJ.Eval} dataset are reported in Table~\ref{tab:basic}.
The Pearson correlation coefficient (PCC) is used to measure the correlation between the assessment score and the human-labelled scores.
It can be observed that the performance of the GOP score outperforms the human-labelled scores ($0.614$ vs. $0.550$),
indicating the robustness of the GOP approach.
We also trained GMM and NF models using the WSJ dataset to compute the PCC between the marginal $p(\mathbf{o})$
and the human-labelled scores. It can be found that the PCCs of both GMM and NF are nearly zero and even negative.
This confirms that $p(\mathbf{o})$ is quite noisy and cannot be directly employed as the assessment score.

\begin{table}[htb!]
 \begin{center}
  \caption{Performance (PCC) on the baseline models.}
  \vspace{-2mm}
   \label{tab:basic}
     \begin{tabular}{l|c|c|c|c}
       \hline
                            &  Human  &   GOP    &   GMM     &   NF     \\
       \hline
             PCC            &  0.550  &   \textbf{0.614}  &   -0.065  &   -0.131    \\
       \hline
     \end{tabular}
 \end{center}
\end{table}

\vspace{-4mm}

\subsection{ASR-free scoring}

This experiment examines the performance of our proposed ASR-free models.
Three marginal models (i-vector, NF and DNF) plus its individual prediction model (SVR) are trained on the \emph{ERJ.Train} dataset.
The results are reported in Table~\ref{tab:free}.
Firstly, it can be seen that the performance of these ASR-free models is inferior to GOP, but is still satisfying.
This indicates that combining the marginal model $p(\mathbf{o})$ and the prediction model $p(s|\mathbf{z})$
is a feasible way to estimate the conditional model $p(s|\mathbf{o})$ and produce the reasonable assessment score.
Secondly, we observed that NF + SVR outperforms i-vector + SVR. The reason is that
NF is a deep generative model, compared with the linear-shallow i-vector model, and can better represent the marginal $p(\mathbf{o})$.
Besides, DNF + SVR obtains the best performance.
This demonstrates that DNF can learn more pronunciation-discriminative representations,
which are more suitable for the downstream assessment prediction.

\begin{table}[htb!]
 \begin{center}
  \caption{Performance (PCC) on the ASR-free models.}
  \vspace{-2mm}
   \label{tab:free}
     \begin{tabular}{l|c|c|c}
       \hline
                       &  i-vector + SVR  &  NF + SVR   &   DNF + SVR   \\
       \hline
             PCC       &    0.434         &  0.441      &   \textbf{0.462}        \\
       \hline
     \end{tabular}
 \end{center}
\end{table}
\vspace{-4mm}

\subsection{Information fusion}

As we mentioned, the posterior $p(\mathbf{q}|\mathbf{o})$ (the GOP score) and the marginal $p(\mathbf{o})$ (the ASR-free score)
are complementary, and could be combined by a score fusion or feature fusion. The results are reported in Table~\ref{tab:fusion}.
The optimal hyper-parameter $\lambda$ was selected based on a small development set which was randomly sampled from the \emph{ERJ.Train} dataset.
Experimental results show that the two fusion approaches consistently achieve better performance than the GOP baseline ($0.614$).
This indicates that these ASR-free approaches can discover some valuable knowledge that has not been represented by GOP,
and also demonstrates that the marginal $p(\mathbf{o})$ involved in the assessment score can relieve the phone-competition problem of GOP.

\begin{table}[htb!]
 \begin{center}
  \caption{Performance (PCC) with information fusion.}
  \vspace{-2mm}
   \label{tab:fusion}
     \begin{tabular}{l|c|c}
       \hline
                            &  Score-fusion                  &  Feature-fusion \\
       \hline
        GOP + i-vector      &    0.640 ($\lambda$ = 0.38)    &     0.625        \\
       \hline
        GOP + NF            &    0.663 ($\lambda$ = 0.34)    &     0.656        \\
       \hline
        GOP + DNF           &    0.676 ($\lambda$ = 0.36)    &     0.667        \\
       \hline
     \end{tabular}
 \end{center}
\end{table}
\vspace{-4mm}

\section{Conclusions}
\label{sec:con}

This paper proposed an ASR-free scoring approach that does not rely on ASR but is based on a generative model.
Our theoretical study shows that this scoring approach offers an interesting correction for the phone-competition problem of GOP,
and empirical study demonstrated that combining the GOP and this ASR-free approach can achieve better performance than the GOP baseline.
Future work will be conducted to understand the behavior of different generative models in the ASR-free assessment,
and study more reasonable fusion approaches to combine the posterior and the marginal.

%\section{Acknowledgements}

\newpage

\bibliographystyle{IEEEtran}

\bibliography{mybib}

\end{document}